\documentclass[3p,times,procedia]{elsarticle}
\usepackage{nupha_ecrc}

\volume{00}

\firstpage{1}

\journalname{Nuclear Physics A}

\runauth{}

\jid{nupha}

\jnltitlelogo{Nuclear Physics A}

\usepackage{amssymb}

\usepackage[figuresright]{rotating}

\begin{document}

\begin{frontmatter}



\dochead{XXVIIth International Conference on Ultrarelativistic Nucleus-Nucleus Collisions\\ (Quark Matter 2018)}

\title{E-by-e jet suppression, anisotropy, medium response and hard-soft tomography}


\author{Y. He$^1$, S. Cao$^2$, W. Chen$^1$, T. Luo$^1$, L.-G. Pang$^{3,4}$ and X.-N. Wang$^{1,3,4}$}\corref{ref1}
\cortext[ref1]{xnwang@lbl.gov}

\address{$^1$Key Laboratory of Quark and Lepton Physics (MOE) and Institute of Particle Physics, Central China Normal University, Wuhan 430079, China \\
$^2$Department of Physics and Astronomy, Wayne State University, Detroit, Michigan 48201, USA\\
$^3$Physics Department, University of California, Berkeley, California 94720, USA\\
$^4$Nuclear Science Division Mailstop 70R0319,  Lawrence Berkeley National Laboratory, Berkeley, CA 94740,USA}

\begin{abstract}
The Linear Boltzmann Transport (LBT) model for jet propagation and interaction in quark-gluon plasma (QGP) has been used to study jet quenching in high-energy heavy-lion collisions. The suppression of single inclusive jet production, medium modification of $\gamma$-jet correlation, jet profiles and fragmentation functions as observed in experiments at Large Hadron Collider (LHC) can be described well by LBT in which jet-induced medium response is shown to play an essential role. In event-by-event simulations of jet quenching within LBT, jet azimuthal anisotropies are found to correlate linearly with the anisotropic flows of bulk hadrons from the underlying hydrodynamic events.
\end{abstract}


\end{frontmatter}


\section{Introduction}
\label{intro}

The strongly coupled quark-gluon plasma (sQGP) created in high-energy heavy-ion collisions at the Relativistic Heavy-Ion Collider (RHIC) and Large Hadron Collider (LHC) has three extreme properties: It is the most perfect fluid with the smallest value of shear viscosity to entropy density ratio $\eta/s\approx (1-2)/4\pi$ \cite{Romatschke:2007mq}; it is most opaque to energetic jets with the jet transport coefficient $\hat q/T^3 \approx 4-8$ \cite{Burke:2013yra}; and it is also the most vortical fluid in nature with a global vorticity $\omega/T\approx 0.001$ \cite{STAR:2017ckg} at the initial temperatures $T\approx 370 - 470$ MeV. These properties of sQGP all involve the collective expansion of the hot QGP medium as governed by relativistic viscous hydrodynamics with fluctuating initial energy density distributions. The initial energy density fluctuation would propagate through the hydrodynamic evolution like sound waves and determine the azimuthal anisotropic flows of the final bulk hadrons that can be used to probe the transport properties of the medium. One interesting question one can ask is what happens to the energetic jets  that propagate through this fluctuating and expanding QGP and how the bulk medium would response to the propagating jets as they deposit energy into the medium. This question is particularly important for the study of reconstructed jets in heavy-ion collisions since the jet-induced medium response will contribute to the energy inside the jet-cone which would become correlated with jets and cannot be subtracted as normal uncorrelated background. Jet quenching on another hand will also depend on the fluctuation in the initial energy density distribution and its azimuthal anisotropy should become correlated with the final bulk hadron anisotropic flows which are determined by the initial energy density fluctuation through hydrodynamic expansion. Such correlations can therefore serve as a hard-soft tomography of the QGP in heavy-ion collisions.

\section{LBT model}
\label{LBT}

The Linear Boltzmann Transport (LBT) model \cite{Li:2010ts,Wang:2013cia,He:2015pra,Cao:2017hhk} has been developed to study jet propagation and modification in dense QGP medium with medium response particularly in mind. It tracks the propagation of the recoil partons from each jet-medium interaction in
the evolving bulk medium as described by a (3+1)D relativistic hydrodynamic model. The interactions between the jet shower partons, recoil partons and thermal partons in the medium are described by the linear Boltzmann equations with both elastic and inelastic scattering matrix elements given by pQCD. 
The inelastic processes in LBT include only induced gluon radiation accompanying each elastic scattering. The radiative gluon spectrum is simulated according to the high-twist approach \cite{Guo:2000nz,Wang:2001ifa}. The probability of elastic and inelastic scattering in each time step are implemented together to ensure unitarity in LBT.

In the Boltzmann transport, initial thermal partons in each scattering are denoted as ``negative'' partons and are allowed to propagate through the medium according to the same Boltzmann equations. They are part of the jet-induced medium excitation (j.i.m.e.) as the diffusion wake behind the propagating jet shower partons \cite{Wang:2013cia,Li:2010ts,He:2015pra}. Their energy and momentum are subtracted from all final observables. The (3+1)D CLVisc  hydrodynamic model~\cite{Pang:2012he,Pang:2018zzo} is used to provide spatial and time information on the local temperature and fluid velocity of the bulk QGP medium during jet propagation. Initial production of jet shower partons per hard nucleon-nucleon collisions is simulated through standard Monte Carlo programs such as Pyhtia. The only parameter in the default version of LBT is the strong coupling constant $\alpha_{\rm s}$ which is found to vary with the colliding energy (0.3 at RHIC and 0.15 at LHC) and is an effective coupling constant for the strong interaction that is regulated by a Debye screening mass. The anti-$k_t$ algorithm is used for jet reconstruction.

\begin{figure}[htbp]
    \centering
  \includegraphics[width=7.2cm]{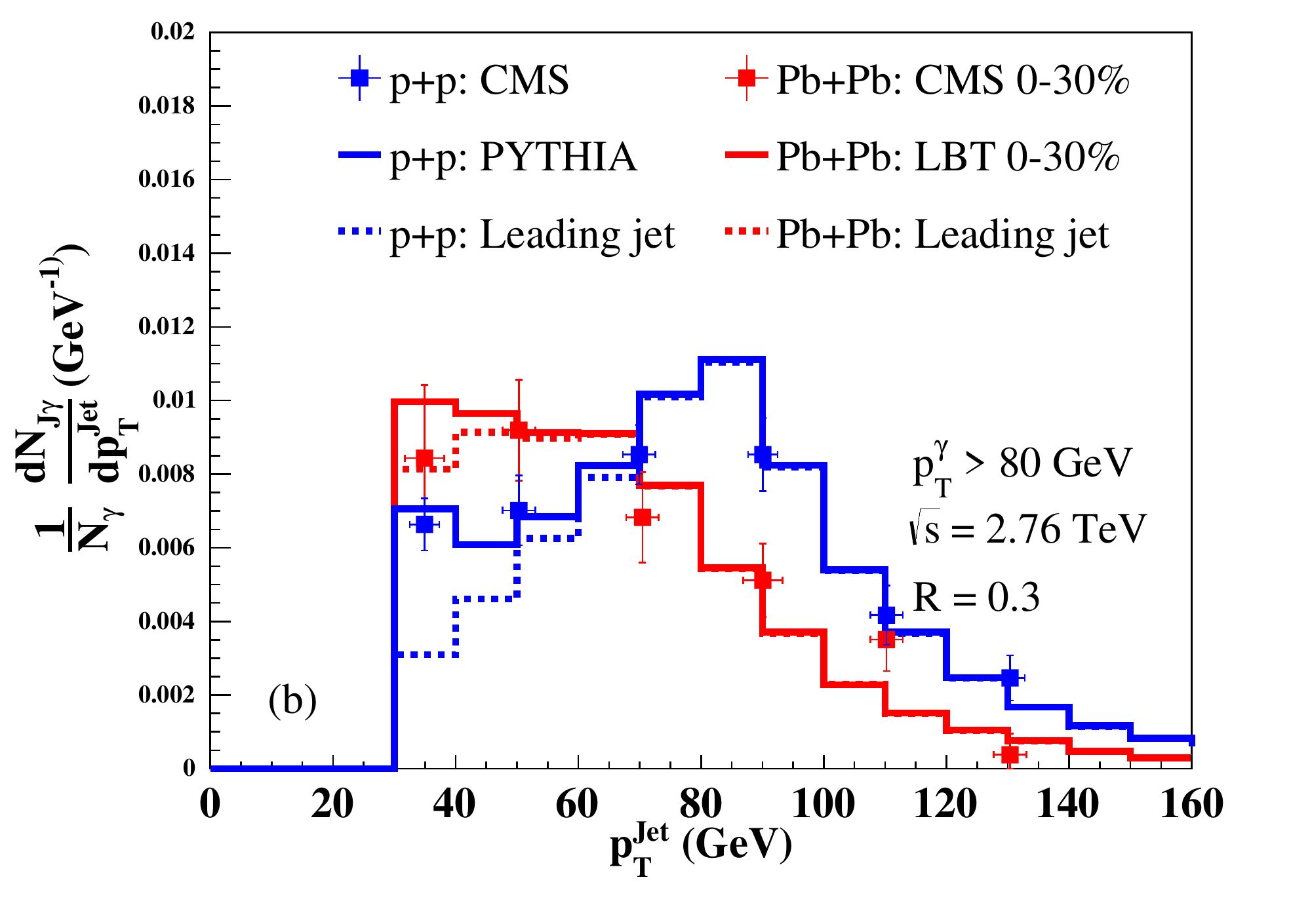}
    \includegraphics[width=7.2cm]{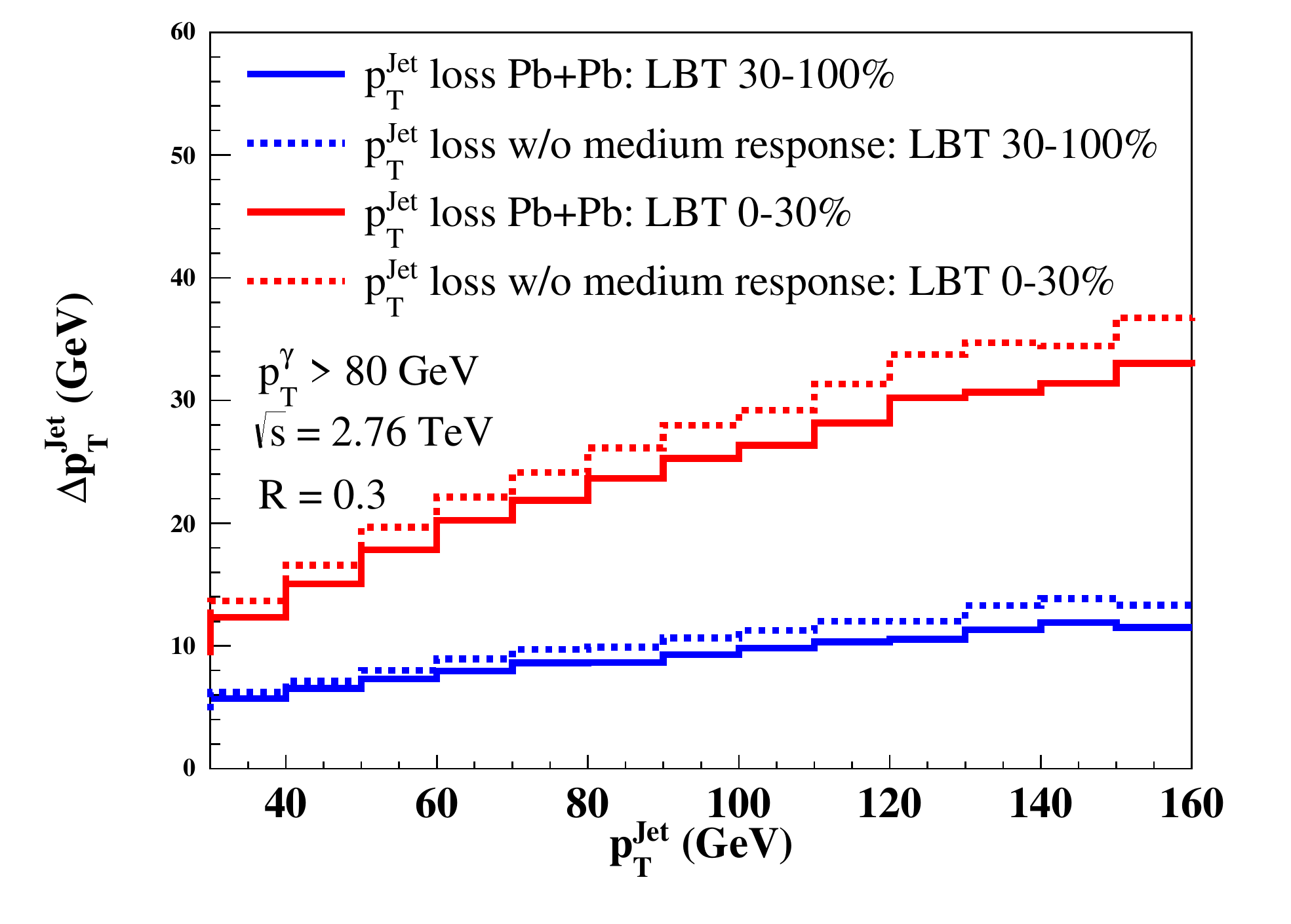}
    \caption{(left) $p_T$ distribution of $\gamma$-jet in central (0--30\%) Pb+Pb (red) and p+p collisions (blue) at $\sqrt{s}=2.76$ TeV from LBT simulations as compared to the CMS experimental data ~\cite{Chatrchyan:2012gt}. (right) Average transverse momentum loss of the leading $\gamma$-jet in Pb+Pb collisions at $\sqrt{s}=2.76$ TeV from LBT as a function of the initial jet $p_T$ with (solid) and without (dashed) contributions from medium response.}
    \label{gammajet}
\end{figure}

\section{Modification of $\gamma$-jets and suppression of single inclusive jets }
\label{gamma-jet}

\begin{figure}[htbp]
    \centering
      \includegraphics[width=5.5cm]{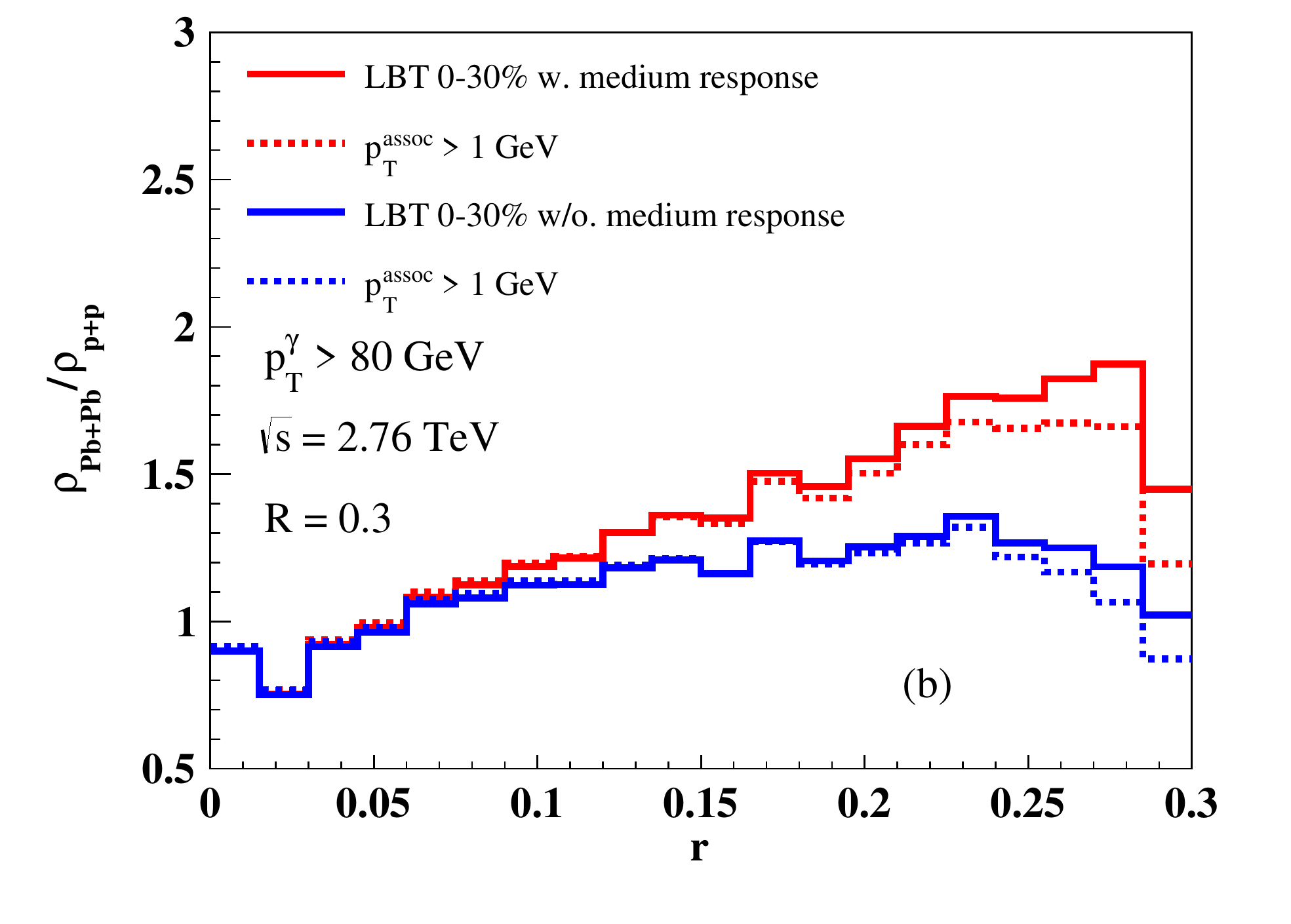}
  \includegraphics[width=9.0cm]{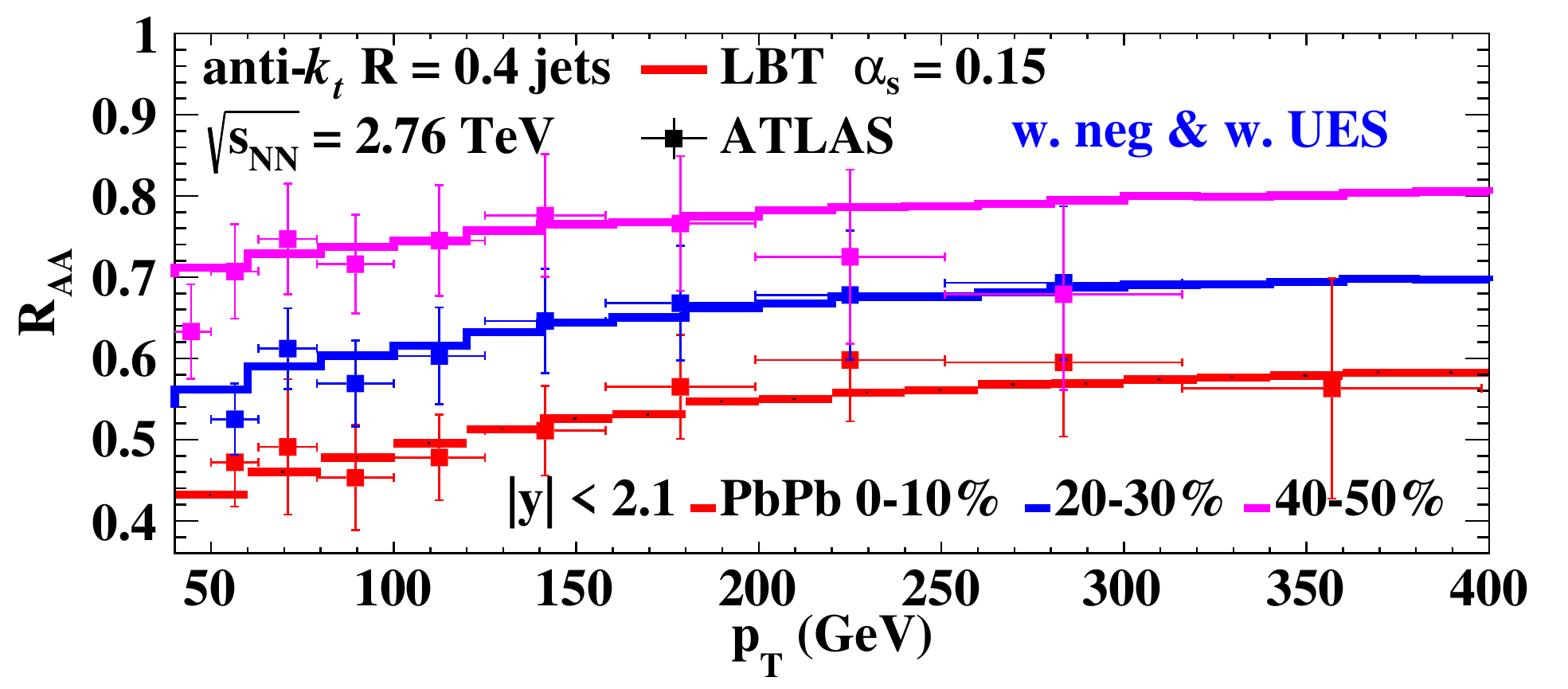}
    \caption{(left) Ratio of $\gamma$-jet profile in 0-30\% central Pb+Pb and p+p collisions at $\sqrt{s}=2.76$ TeV.  (right) Suppression factor
    $R_{AA}$ for single inclusive jet production in 0-10\%, 20-30\% and 40-50\% Pb+Pb collisions at $\sqrt{s}=2.76$ TeV as compared to ATLAS data \cite{Aad:2014bxa}.}
    \label{singlejet}
\end{figure}

LBT model has been employed successfully to describe light and heavy flavor hadron suppression \cite{Cao:2017hhk}, $\gamma$-jet modification \cite{Wang:2013cia,Luo:2018pto} and single inclusive jet suppression in heavy-ion collisions. Shown in Fig.~\ref{gammajet}(left) is the $p_T$ distribution of $\gamma$-triggered jets in central Pb+Pb collisions from LBT as compared to CMS data \cite{Chatrchyan:2012gt}. Both LBT results and the data show clearly a shift of the distribution due to jet energy loss. One can calculate this jet energy loss within LBT mode which is defined as the difference between the final jet energy in heavy-ion collisions and the corresponding jet energy in p+p collisions from the same initial jet shower partons.
As shown in Fig.~\ref{gammajet}(right), the average jet energy loss increases logarithmically with the initial jet energy and the inclusion of medium response reduces noticeably the energy loss. The inclusion of medium response in the jet reconstruction can reduce the final jet energy loss as much as 30\% for jets with initial jet energy about 400 GeV. The medium response also enhances considerably the jet profile at large radius \cite{Luo:2018pto} as shown in Fig.~\ref{singlejet}(left).

\begin{figure}[htbp]
    \centering
   \includegraphics[width=7.00cm]{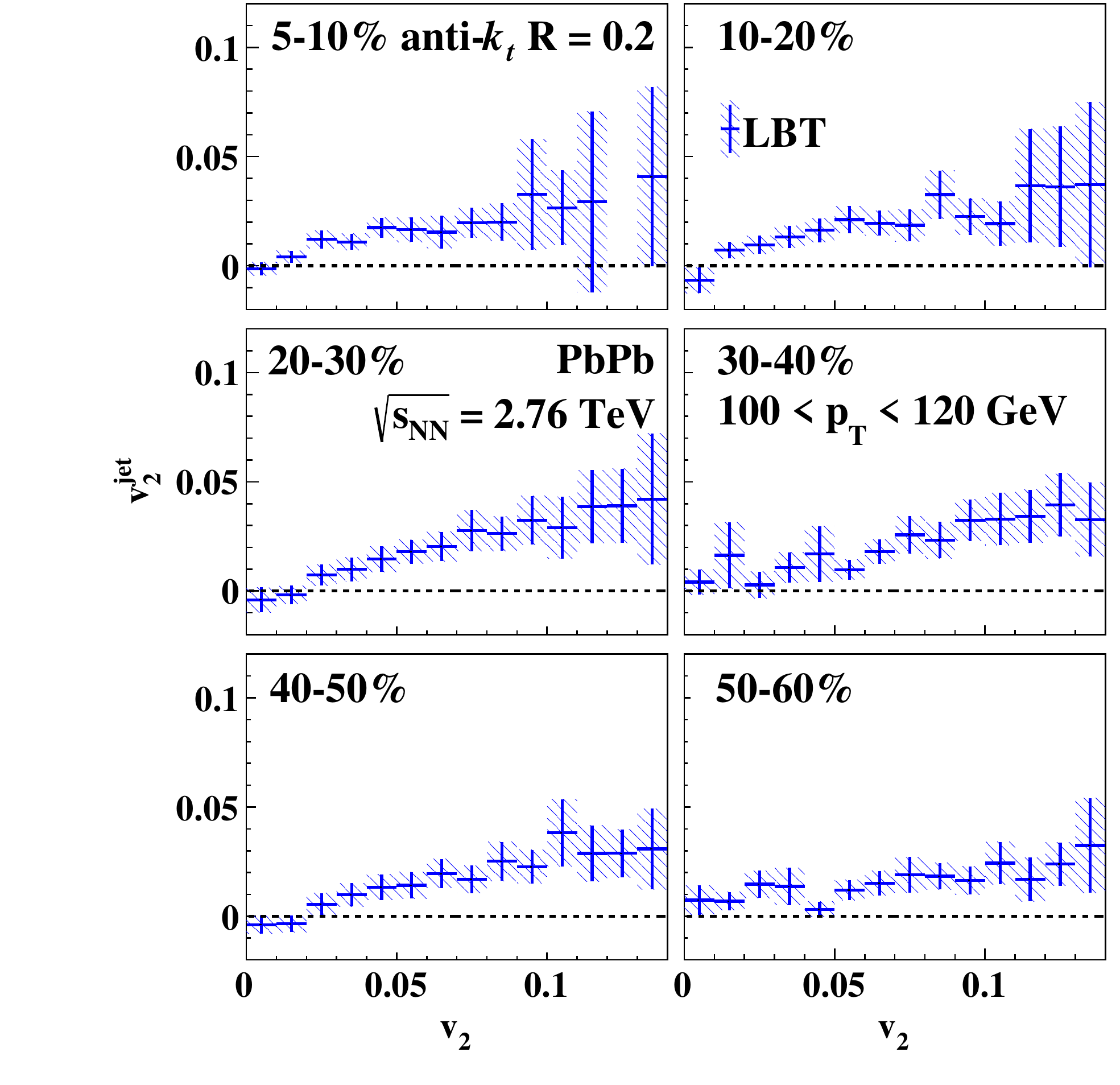}\hspace{0.5cm}
      \includegraphics[width=5.80cm]{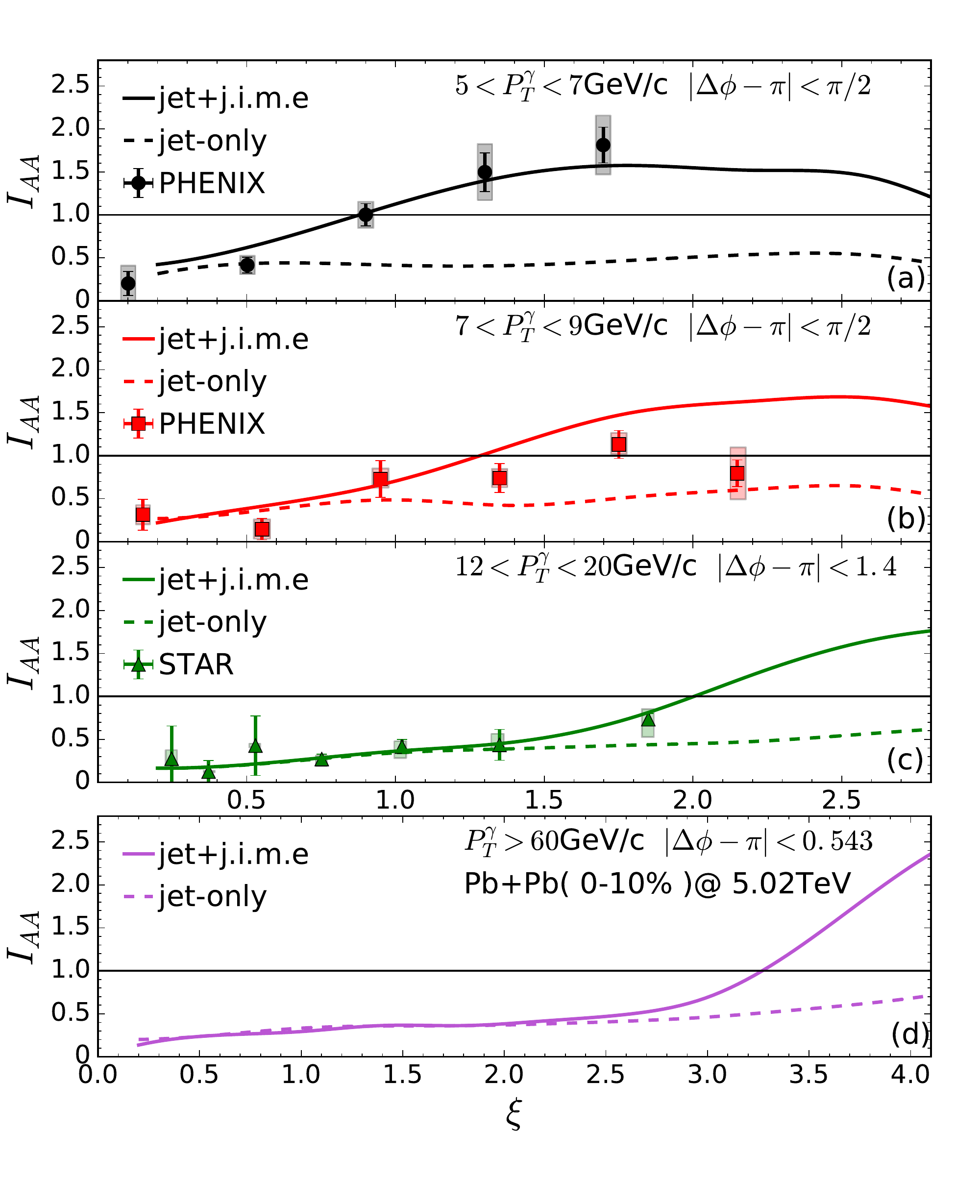}
    \caption{(left) Azimuthal anisotropy $v_2^{\rm jet}$ of hard jets as a function of the elliptic flow of bulk hadron spectra from the underlying hydro events as given by e-by-e LBT simulations in Pb+Pb collisions at $\sqrt{s}=2.76$ TeV with different centralities.  (right) CoLBT-hydro results on modification factor for $\gamma$-hadron correlation as a function of $\xi=\log(p_T^\gamma/p_T^h)$ in Au+Au collisions at RHIC and Pb+Pb collisions at LHC as compared to data from PHENIX \cite{Ge:2017irb} and STAR \cite{STAR:2016jdz}.}
    \label{gammahadron}
\end{figure}

To study the hard-soft correlation due to event-by-event (e-by-e) fluctuation of the QGP medium, we have carried out e-by-e simulations of jet quenching in Pb+Pb collisions at LHC with 200 fluctuating hydro profiles for each centrality. The hydro profiles are generated with CLVisc model \cite{Pang:2012he,Pang:2018zzo} with fluctuating initial conditions from the AMPT model \cite{Lin:2004en}. As shown in Fig.~\ref{singlejet}(right), e-by-e simulations from LBT can describe well the averaged suppression factor of single inclusive jet in heavy-ion collisions with different centralities. In these e-by-e simulations, one can also calculate the jet azimuthal anisotropy $v_2^{\rm jet}$ which has a clear linear correlation with the elliptic flow of the final bulk hadron spectra from the underlying hydro events as shown in Fig.~\ref{gammahadron} (left). Such hard-soft correlation can be employed to study the initial energy density fluctuation and transport properties of the bulk QGP medium.

\section{CoLBT-hydro}
\label{colbt}

Under the linear approximation ($\delta f\ll f$) in LBT, interaction among jet shower and recoil partons is neglected. This assumption will break down when the jet-induced medium excitation becomes comparable to the local thermal parton density.   To go beyond this region of applicability, a coupled LBT and hydrodynamic (CoLBT-hydro) model \cite{Chen:2017zte} is developed in which jet transport in LBT is coupled to the hydrodynamic evolution of the bulk medium in real time. The coupling is through a source term in the hydrodynamic equations determined by the energy and momentum deposited by the propagating jet shower partons. Shown in Fig.~\ref{gammahadron} (right) are medium modification factors for $\gamma$-hadron correlation as a function of $\xi=\log(p_T^\gamma/p_T^h)$ from CoLBT-hydro as compared to experimental data from PHENIX \cite{Ge:2017irb} and STAR \cite{STAR:2016jdz}. As one can see, jet quenching leads to the suppression of leading hadrons from the fragmentation of $\gamma$-jets at small $\xi$. At large $\xi$, jet-induced medium excitation (j.i.m.e.), however, leads to an enhancement of soft hadrons. The onset of the enhancement is found to occur at a fixed value of $p_T^h\approx 2$ GeV. We have also calculated the modification of $\gamma$-jet fragmentation function and will report the results in a separate publication.

\section{Summary}
\label{summary}

In summary,  $\gamma$-jet, $\gamma$-hadron correlations, suppression of single inclusive jets and jet azimuthal anisotropy have been studied with the LBT model and CoLBT-hydro for jet transport in heavy-ion collisions. The effects of jet-induced medium response are found important for the description of jet energy loss, modification of jet profiles and fragmentation functions. The azimuthal anisotropy of jet quenching is found to correlate well with the geometry fluctuation of the initial energy density and the bulk hadron anisotropic flows. 

\section*{Acknowledgement}

This work is supported by the NSFC under grant No. 11521064, NSF within the JETSCAPE Collaboration and 
 under grant No. ACI-1550228, and ACI-1550300, and U.S. DOE under Contract Nos. DE-AC02-05CH11231 and DE-SC0013460. 
 Computations are performed at the Green Cube at GSI, GPU workstations at CCNU and the DOE NERSC.
 







\end{document}